\documentclass[12pt,preprint]{emulateapj}

\usepackage{amsmath}

\begin{document}

\title{Population Synthesis of Galactic Be-star Binaries with A Helium-star Companion
}
\author{
Yong Shao$^{1,2}$ and Xiang-Dong Li$^{1,2}$}

\affil{$^{1}$Department of Astronomy, Nanjing University, Nanjing 210046, People's Republic of China; shaoyong@nju.edu.cn}

\affil{$^{2}$Key laboratory of Modern Astronomy and Astrophysics (Nanjing University), Ministry of
Education, Nanjing 210046, People's Republic of China; lixd@nju.edu.cn}

\begin{abstract}

LB-1 was originally suggested to harbour a very massive ($ \sim 70 M_\odot $) black hole, 
but was recently suggested to be a post-mass transfer binary containing a Be star and a helium (He) star. 
In this paper, we use the binary population synthesis method to simulate 
the potential population of the Be$ - $He binaries in the Milky Way. 
Mass transfer process during the progenitor binary evolution plays a vital role in determining the 
possible properties of  the Be$ - $He binary population. By constructing a range of physical models with significantly 
different mass-transfer efficiencies, we obtain the predicted distributions 
at the current epoch of the component masses and the orbital periods 
for the Be$ - $He binaries. In particular, we show that, LB-1 very likely has evolved through non-conservative 
mass transfer if it is indeed a Be$ - $He system. We estimate that there are more than $ 10^{3} $ Be$ - $He
binaries with V-band apparent magnitudes brighter than LB-1.

\end{abstract}


\section{Introduction}

\citet{liu19} reported the discovery of a unique binary system LB-1 that may harbour a $ \sim 70 M_\odot $ 
black hole with a B-type star in a $ \sim79 $ day orbit. The formation of such a massive black hole 
would drastically challenge current stellar evolution theories, since it is not expected to be created
in a Galactic high-metallicity environment due to strong stellar winds \citep[e.g.,][]{bbf10}
and pair-instability supernovae \citep[e.g.,][]{fr19}. 
This has led to some exotic scenarios
for the formation of this source \citep{gf19,bh20}. On the other hand, other groups \citep[e.g.,][]{am20,eq20,es20,ig20,sm20} 
suggest that LB-1 may actually host a normal stellar-mass black hole. Until now, the nature of LB-1 is still under debate \citep{liu20}.
More recently, \citet{sb20} argued that LB-1 does not contain a compact star but instead a post-mass transfer binary 
comprising of a $ \sim7M_\odot $ Be star and a $ \sim1.5M_\odot $ stripped star. 

The formation of Be stars is not fully understood, but there is a consensus that they are rapid rotators with the equatorial
velocities near their Keplerian limits \citep[e.g.,][]{pr03}. It has been proposed that three main mechanisms are responsible for
the rapid rotation of Be stars:
(1) they were born as rapid rotators by inheriting the angular momentum from their parental molecular clouds \citep[see however,][]{hg10};
(2) angular momentum transfer from the core to the envelope during the main-sequence evolution \citep[e.g.,][]{em08,hwl20}; 
(3) they are the mass gainers or the stellar mergers as a consequence of binary interaction \citep[e.g.,][]{rv82,pc91,sl14}.
Investigations on the observed sample of Be stars suggest that a large majority of them have formed via binary mass transfer \citep{mg05,bss20}.

Observations show that massive stars of spectral types O and B are born predominately in binary
and multiple systems \citep{sd12,kk14,md17}. The majority of them will interact with a
binary companion via mass transfer, which significantly changes the evolutionary paths 
of both stars and the orbital parameters of the binary system. A fraction of binary systems are expected to evolve through
a stable Roche-lobe overflow phase, during which the primary star loses most of its hydrogen envelope 
while the secondary star is rejuvenated and spun up due to mass accretion \citep{h02}. After the mass transfer, the primary
evolves to be a stripped helium (He) star (also referred as hot subdwarf star or Wolf-Rayet star) and the secondary appears to be a rapidly 
rotating Be star, leading to the formation of a Be$-$He binary \citep{pc91,sl14,lb20}. 
The subsequent evolution of such systems is likely to become Be/X-ray binaries if the He stars are massive enough to 
quickly evolve into compact stars. 
Population synthesis studies indicate that the number of Galactic Be$-$He binaries is of the order $ 10^{5} $,
which is comparable with that of Be$-$white-dwarf systems and about two orders of magnitude more than that of both 
Be$-$neutron-star and Be$ - $black-hole systems \citep{sl14}. In this paper, we focus on the detailed properties of 
Galactic Be$-$He binaries and the formation of the peculiar system LB-1.

To date, there are several Be$-$He systems
confirmed in the Milky Way \citep{wg18}, e.g. $ \varphi $ Persei \citep{mm15}, FY CMa \citep{pg08}, 
59 Cyg \citep{pp13}, 60 Cyg \citep{wg17}, HR 2142 \citep{pw16}, and $\textit{O}$ Puppis \citep{kk12}. 
More recently, HR 6819 was also suggested to be a Be$-$He binary \citep{bsm20} rather than 
a triple system with a black hole \citep{rb20}.
The orbital periods of these Be$-$He binaries vary in the range
of $ \sim 30-150 $ days. All in the observed sample host a He star with mass around $ 1M_\odot $.
The masses of the Be star are distributed in the range of  $ \sim 6-12M_\odot$. 

One of the most important parameters that can significantly influence the properties of Be$-$He systems is the 
mass-transfer efficiency (i.e. the fraction of the accreted matter by the secondary star among all the transferred matter). 
Based on the parameters of the well-studied source $ \varphi $ Persei, \citet{sg18} evolved an extensive grid of binary systems to 
reproduce this binary by varying the mass-transfer efficiency in a range of $ 0.25-1.0 $. They concluded that this system 
must have experienced a near-conservative mass transfer stage, which is consistent with earlier studies of \citet{vd98} and \citet{p07}.
It has been suggested that, however, the evolution of massive binaries can also experience significant mass loss during the mass-transfer
process \citep[e.g.,][]{dm07,sl16}. 
  
LB-1 is located at a distance of $ \sim2-4 $ kpc and has a V-band apparent
magnitude of $ \sim 12 $ mag \citep{liu19}, which is much fainter than
other observed Be$-$He binaries. Whether LB-1 is a black hole binary or a Be star binary will considerably influence  the strategy of black hole search with optical observations. It is also noted that, although there are a few theoretical investigations on the formation of  
specific Be$ - $He binaries, a systematic work on the whole population is still lacking.
In this paper, we explore the possible properties of the Be$-$He systems in the Milky Way. 
Our main goal is to obtain the predicted number and the parameter distribution of the Be$-$He binary population,
by taking into account different mass-transfer models (efficiencies) during the progenitor binary evolution.
The remainder of this paper is organized as follows.  In Section 2, we present the binary population synthesis 
\citep[BPS, see][for a review]{han20}
method. We present the calculated results in Section 3.  We briefly discuss their implications
in Section 4 and conclude in Section 5.

\section{Method}

We employ the \textit{BSE} code originally developed by \citet{h02} to deal with the formation and evolution of 
Be$-$He binaries. Significant modification in this code has been made by \citet{sl14} to deal with the mass-transfer 
stability and common envelope evolution. 
This code makes use of a series of fitting formulae \citep{h00} to describe the structure of 
single stars as they evolve as a function of the stellar mass and age, so it can rapidly compute the evolution 
of millions of binary systems. Modelling the evolution of a binary system involves detailed treatments of stellar
winds, tidal interactions, and mass and angular momentum transfer. 

Starting from a primordial binary with two zero-age 
main sequence stars, the primary star first evolves to fill its Roche lobe and supplies its
envelope material to the secondary star. Mass accretion onto the secondary star causes it to 
expand and spin up. The expanded size of the secondary star is strongly dependent on the mass 
accretion rate \citep[e.g.,][]{n77}.  Usually rapid 
mass accretion can drive the secondary star to get out of thermal equilibrium and significantly expand,  
leading to the formation of a contact binary if the secondary star has also filled its own Roche lobe \citep{ne01}. 
Following \citet{sl14} we build three mass transfer models. In Model I, the mass accretion rate 
onto the secondary is assumed to be the mass-transfer
rate multiplied by a factor of $ (1- \Omega / \Omega_{\rm cr}$), where $ \Omega $ is the angular velocity of the 
secondary star and $ \Omega_{\rm cr} $ is its critical value \citep[see also][]{se09}. By simulating the evolution of massive 
binaries under the assumption of this model, \citet{sl16} pointed out that the averaged mass-transfer efficiencies decrease from
$ \sim 0.7 $ to $ \sim 0.1 $ with increasing binary orbital periods.
%in wide systems is highly non-conservative with an efficiency as low as $ \lesssim 0.2 $. 
In Model II, it is
assumed that the secondary star accretes half of the transferred matter with a constant mass-transfer efficiency of 0.5. 
In Model III, the mass accretion rate is 
assumed to be limited by a factor of $ \min [\rm 10 (\tau_{\dot{M}}/\tau_{KH}), 1 ]$, where 
$\rm \tau_{\dot{M}} $ is the mass-transfer timescale and $\rm \tau_{KH} $ is the thermal timescale of 
the secondary star \citep{h02}. Generally the mass-transfer process in Model III is near-conservative \citep{sl14}, 
and we demonstrate that the averaged mass-transfer efficiencies for almost all 
binaries are higher than 0.95 after dealing with the data of our BPS calculations.
In all the cases, the escaped material from the binary system is assumed to carry away the specific orbital 
angular momentum of the secondary star. 
As a consequence, the maximal initial mass ratios of the primary to the secondary star
for avoiding the contact phase can reach as high as
$ \sim 6 $ in Model I, and drops to $ \sim 2.5$ and $ \sim2.2 $ in Model II and III, respectively \citep{sl14}.
We assume that all contact binaries evolve into a common-envelope phase.

We include the effect of stellar winds, tides and mass exchange on the rotational velocity of the 
secondary star \citep[see e.g.,][]{dm13}. The internal rotational profile
of the secondary star is approximated to be rigid rotation, which is a reasonable approximation for the secondary
star residing at the main-sequence stage \citep[e.g.,][]{mm05,bd11}. When the mass transfer occurs via Roche lobe overflow, 
the rotational evolution of the secondary star is 
mainly controlled by the competition between the spin-down effect due to tidal interactions 
and the spin-up effect due to mass and angular momentum accretion.  
We follow \citet{h02} to treat the spin-orbit coupling between binary components due to tidal interactions. 
During the mass-transfer process, the transferring matter may either form an accretion disk around the secondary star
or directly impact on its surface, depending on the minimum distance $ R_{\rm min} $  between the mass stream and
the secondary star \citep{ls75} compared with the radius $ R_{\rm s} $ of the secondary star. 
If $ R_{\rm min} > R_{\rm s} $, the mass stream is assumed to collide with itself after which the viscous process 
leads to the formation of an accretion disk. It is assumed that the secondary star
accretes matter from the inner edge of the disk with the specific angular momentum of  $ \sqrt{G M_{\rm s} R_{\rm s}} $, 
where $ G $ is the gravitational constant and $ M_{\rm s} $ is the secondary mass.
Otherwise, the matter stream impacts directly on the surface of the secondary star and the specific angular 
momentum of the impact stream is estimated to be $ \sqrt{1.7 G M_{\rm s} R_{\rm min}} $.

According to the observational characteristics of Be stars,
%we use a phenomenological definition to identify them, that is,  
we identify them as the main-sequence stars with masses of $ 3-23M_\odot $ 
(spectral types of A0$ - $O9) and rotational velocities
exceeding 80\% of their Keplerian limits \citep{s82,n98,pr03}. As shown by \citet{sl14}, the formation
of Be stars in binary systems does not favour a common-envelope phase during the primordial evolution, 
so we only consider the case of stable mass transfer via Roche lobe overflow to produce Be stars in our calculations. 
%The He stars in the present context refer to stripped-down, bare He-burning stellar cores.
Evolved from the primordial binaries, the formation of Be$ - $He binaries usually experiences 
either Case A or Case B mass transfer \citep{kw67,p67}. 
For relatively close primordial binaries, mass transfer starts when the primary stars are still on the main-sequence
phase with a burning hydrogen core (Case A mass transfer). The primary stars will experience a rapid 
contraction after the depletion of fuel in their convective cores, leading to a temporary detachment of the binary systems. 
When the primary stars expand due to shell hydrogen burning and fill their
Roche lobes again, Case AB mass transfer takes place. For relatively wide
primordial binaries, the primary stars are on the shell hydrogen burning phase when mass transfer begins (Case B mass transfer). 
We follow \citet{h02} to age the primary stars and rejuvenate the secondary stars due to mass exchange. 
After mass transfer, the primary stars leave stripped-down, bare He-burning stellar cores that are referred to He stars in 
the present context.

\begin{figure*}[hbtp]
\centering
\includegraphics[width=1.0\textwidth]{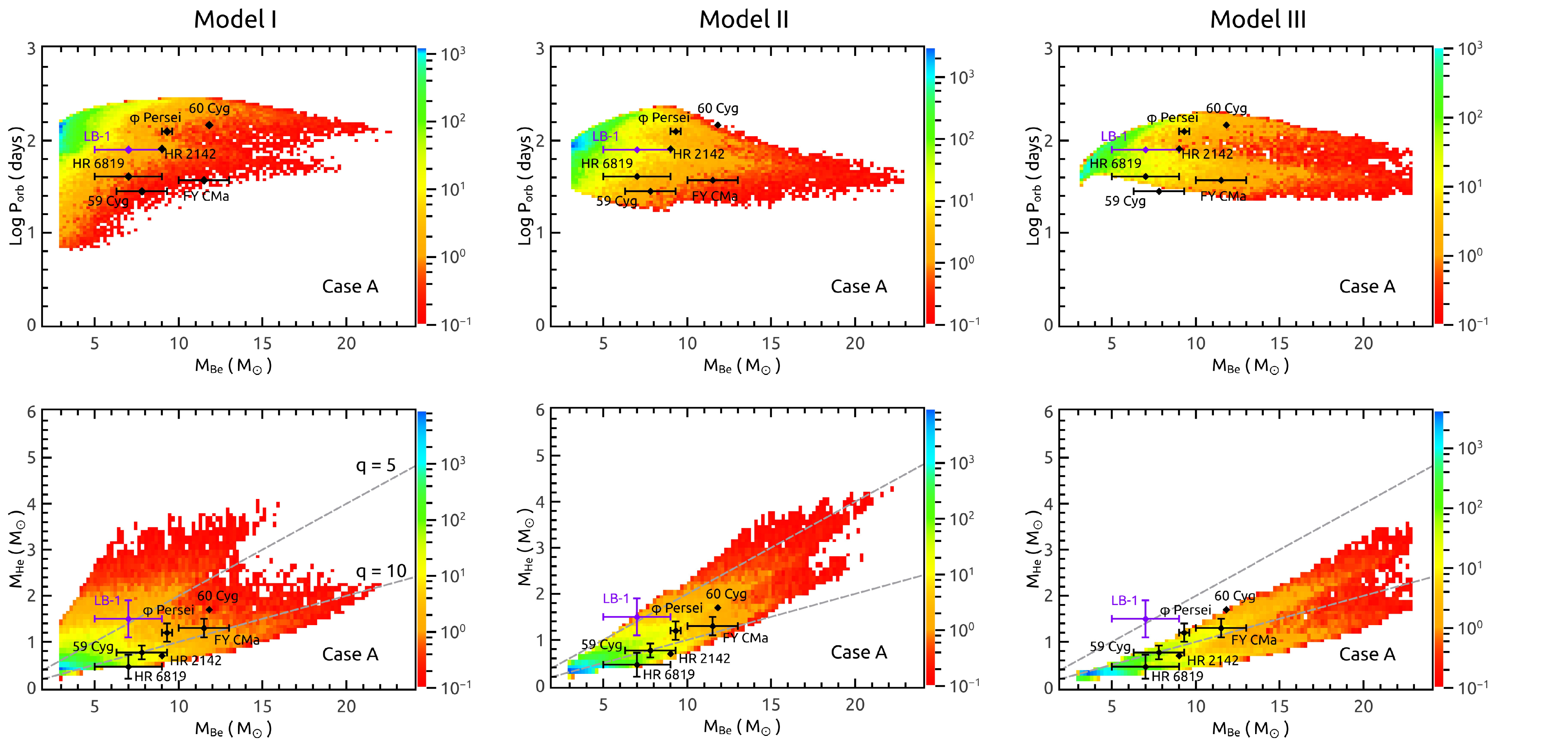}
\includegraphics[width=1.0\textwidth]{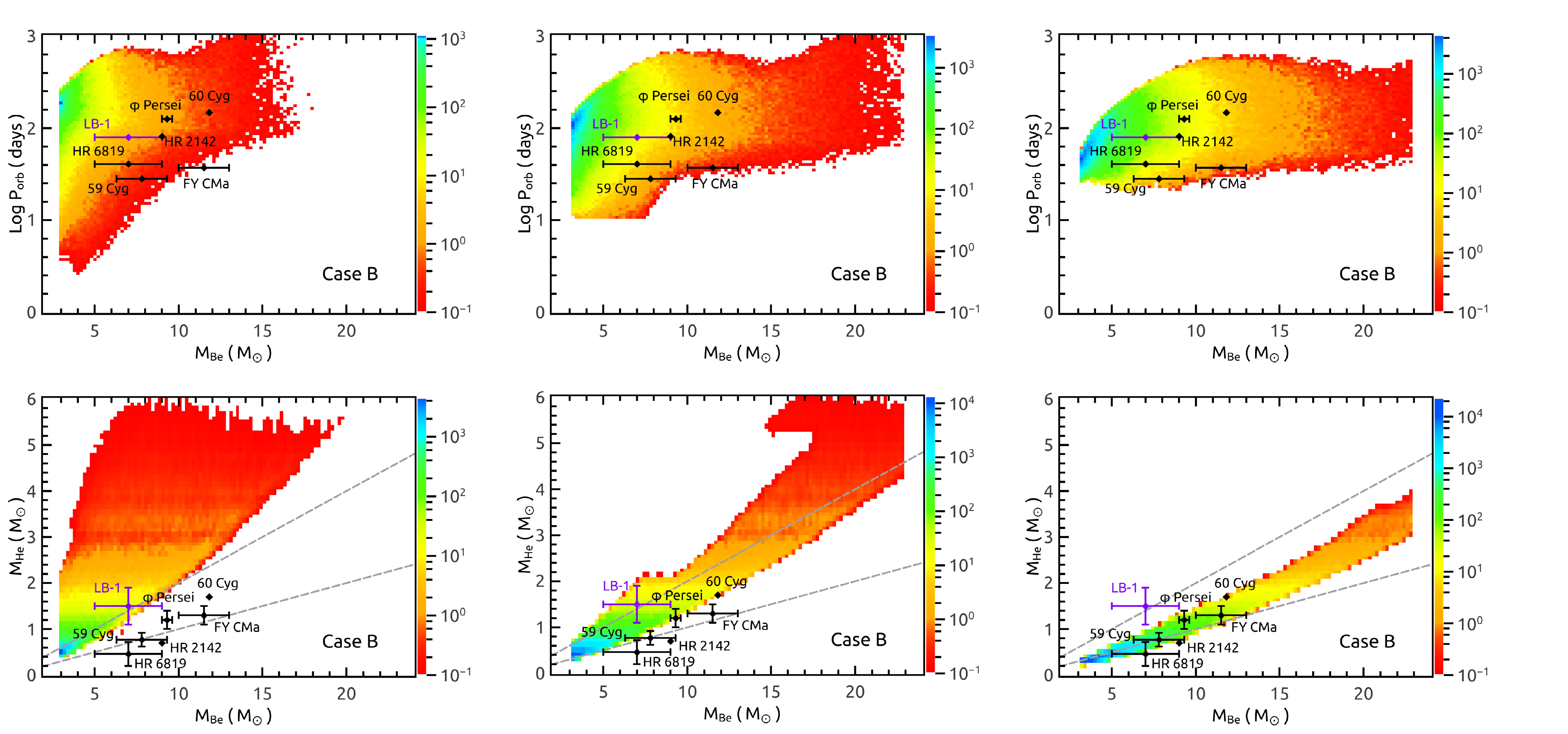}
%\linespread{0.7}
\caption{Calculated number distributions of Galactic Be$-$He binaries in the $ M_{\rm Be}-P_{\rm orb} $
and $ M_{\rm Be}-M_{\rm He} $  planes under the assumption of Models I$ - $III (from left to right). 
The top and bottom six panels correspond to the evolution of the progenitor systems experienced the process of 
Case A and Case B mass transfer, respectively.
The positions of suggested Be$-$He systems are plotted with solid diamonds$-$the purple one corresponds to 
LB-1 while the black ones correspond to other binaries.  For 60 Cyg and HR~2142 without
error bars in mass, the Be-star masses are crudely estimated from their spectral classifications \citep{kh00,pw16}.
The colors in each pixel are scaled according to the corresponding numbers, only the systems with numbers larger than 0.1 are
plotted.
The two dashed lines correspond to mass ratios $ q$  ($= M_{\rm Be}/M_{\rm He} $) equal to 5 and 10 for comparison. 
   \label{figure1}}
\end{figure*}

We have evolved a large number ($ \sim 10^{7} $) of binary systems in each model, by setting a grid of initial parameters 
as follows. The primary masses $ M_{\rm 1} $ vary in the range of $ 1-60 M_\odot $, the secondary masses $ M_{\rm 2} $ 
in the range of $ 0.1-60 M_\odot $, and the orbital separations $ a $ in the range of $ 3-10^4 R_\odot $. 
For the primary stars, less massive ($ \lesssim 2M_\odot $) ones can only develop degenerate He 
cores \citep{h00}, and more massive ($ > 60 M_\odot $) ones are extremely rare due to the initial mass function (IMF). 
For the secondary stars, only the systems with $ M_{2} < M_{1} $ are evolved. 
Each parameter $ \chi $ is logarithmically spaced with the $ n_{\chi} $ grid points, thus 
\begin{equation}
\delta \ln \chi = \frac{1}{n_{\chi}-1} (\ln \chi_{\rm max} - \ln \chi_{\rm min}).
\end{equation}
If a specific binary $ i $ evolves through a phase that is identified as a Be$-$He binary, then the system can 
contribute the population with a rate
\begin{equation}
\begin{aligned}
R_{i} =  \left( \frac{f_{\rm b}}{f_{\rm s}+ 2 f_{\rm b}} \right) \left( \frac{S}{M_{*}} \right) \Phi (\ln M_{1i}) \varphi (\ln M_{2i})
\Psi (\ln a_{i}) \\ 
\delta \ln M_{1} \delta \ln  M_{2} \delta \ln  a,
\end{aligned}
\end{equation}
where $ f_{\rm s} $ and $ f_{\rm b} $ are respectively the fractions of single stars and binary (including multiple) systems 
among all stars,
so the relation of $ f_{\rm s} + f_{\rm b} = 1$ is always satisfied. Here $ S $ is the star formation rate of the Milky Way and 
$M_{*} \sim 0.5 M_{\odot}$ is the average mass of all stars.
Considering the star formation history of the Milky Way, we use a constant 
star formation rate of $ S = 3 M_{\odot}\,\rm yr^{-1} $ over the past 10 Gyr period \citep{sb78,dh06,rw10}. 
$ \Phi (\ln M_{1i}) $, $\varphi (\ln M_{2i})$ and $ \Psi (\ln a_{i}) $ are the normalized functions to weight
the contribution of the specific binary with logarithmic parameters of  $\ln M_{1i} $, $\ln  M_{2i} $ and $\ln a_{i} $, respectively.
Since the initial parameters of primordial binaries cover rather 
wide ranges, all input values and the corresponding distribution functions are taken in the logarithmic forms.
A more detailed description can be found in the method section of \citet{h02}.
The masses of the primary stars follow the \citet{k93} IMF,
\begin{equation}
  \xi (M_{1} ) = \left\{
    \begin{array}{ll}
      0                              & M_{1} \leq 0.1M_\odot \\
      a_{1} M_{1}^{-1.3}  & 0.1M_\odot < M_{1} \leq 0.5M_\odot \\
     a_{2} M_{1}^{-2.2}  & 0.5M_\odot < M_{1} \leq1.0M_\odot  \\
     a_{2} M_{1}^{-2.7}   & 1.0M_\odot <  M_{1} <  \infty.
    \end{array}  \right.,
\end{equation}
where $ a_{1} = 0.29056$ and $ a_{2} = 0.15571$ are the normalized parameters. Then 
\begin{equation}
\Phi (\ln M_{1}) = M_{1}  \xi (M_{1} ).
\end{equation}
We assume that the secondary masses $M_{2}$ obey a flat distribution between 0 and $ M_{1}$ \citep{kf07}, 
thus giving
\begin{equation}
\varphi (\ln M_{2}) = \frac{M_{2}}{M_{1}}.
\end{equation}
The distribution of the initial orbital separations $ a $ is assumed to be logarithmically uniform 
between $ 3$  and $10^4 R_\odot $ \citep{a83}, thus we obtain
\begin{equation}
\Psi (\ln a)  = k = \rm const.
\end{equation}
The normalization of this distribution gives $ k = 0.12328 $.
%We simulate the evolution of the primordial binaries by setting the initial parameters as follows. 
%The masses of the primary stars follow the initial mass function (IMF) suggested by \citet{k93}, 
%which are set to be in the range of $ 1-60 M_\odot $ drawn from a power-law 
%distribution with index $ -2.7 $. Less massive ($ \lesssim 2M_\odot $) stars can only develop degenerate He 
%cores \citep{h00}, and more massive ($ > 60 M_\odot $) stars are extremely rare due to the IMF. 
%We assume that the mass ratios of the secondary to the primary star obey a flat distribution between 0 and 1.
%The distribution of the initial orbital separations is assumed to be logarithmically uniform 
%in the range of $ 3-10^4 R_\odot $ \citep{a83}. 
We assume that all binaries initially have circular orbits, as the outcome of the interactions of 
systems with the same semilatus rectum is almost independent of eccentricity \citep{h02}. 
All stars are assumed to be initially in binaries\footnote{Observations show that most ($f_{\rm b} \sim 0.6-0.9  $) of
OB stars are born as members of binary and multiple systems \citep{md17}, this can reduce our calculated birthrate
and expected number of Galactic Be$-$He binaries by a factor of less than 2.}.
The initial metallicity of stars 
is taken to be $ Z = 0.02 $. The Galactic thin disk is an active site for ongoing star formation, 
which dominates the formation of relatively massive binaries that we are working with. 
Observations of HII regions via radio continuum emission indicate that ongoing star
formation occurs with a nearly uniform efficiency over the Galactic thin disk \citep[][and references therein]{ke12}.
We simply assume that the Galactic thin disk is infinitesimally thin and the recent star-formation 
rate has a uniform distribution over the disk.
We adopt the distance of the Sun with respect to the Galactic center to be 8 kpc \citep{fw97}. 
The Be$-$He binaries are assumed to locate at their birth places without considerable movement, 
considering the relatively small velocities and short lifetimes of massive binaries.  
%We follow the method of \citet{sl19} to deal with 
%the apparent magnitudes of binary components in the V band, taking into account the effect of interstellar extinction. 

For the detection of Be$-$He binaries, we can predict the possible distributions of some observational parameters  
including the apparent magnitude and the radial velocity semi-amplitude of the binary components. 
Based on stellar luminosity $ L $, effective temperature $ T_{\rm eff} $ and surface gravity $ g $, we edit a subroutine 
in the code to yield the V-band absolute magnitude $ M_{V} $.  Under the scheme of  Vega magnitude system, \citet{gbb02} 
tabulated the bolometric corrections with a grid of $ T_{\rm eff} $ and $ \log\, g $, which were obtained from the 
combination of the real atmospheric spectra \citep{f94,c97,a00} and the Planck black-body spectra (for very hot stars). 
In our calculations, the bolometric correction
for a specific star is given by a linear interpolation in the existing grid.
By considering the interstellar extinction $ A_{\rm V} $ in the V band, 
the apparent magnitude $ m_{\rm V} $ can be given as
\begin{equation}
m_{\rm V} = M_{\rm V} + 5\left( 2+\log \left( \frac{D}{1\,\rm kpc}\right) \right)  + A_{\rm V}(D), 
\end{equation}
where $ D $ is the distance of the binary system from the Sun.
Since the average extinction
of the Galactic disk is $ \sim 1 \rm \, mag \, kpc^{-1} $ \citep{s78}, we adopt $A_{\rm V}(D) = \frac{D}{1\,\rm kpc}$. 
To estimate the radial velocity semi-amplitude $ K $, the orbital inclination 
of Be$-$He binaries is assumed to distribute uniformly between 0 and $ 2 \pi $.
From the BPS outcomes, we select all Be$-$He binaries and record relevant parameters at each of the
evolutionary steps. The corresponding number of a specific type of binary can be evaluated by multiplying its birthrate
with the timestep. 
%in which the effect of interstellar extinction has been taken into account. 

\begin{figure}[hbtp]
\centering
\includegraphics[width=0.4\textwidth]{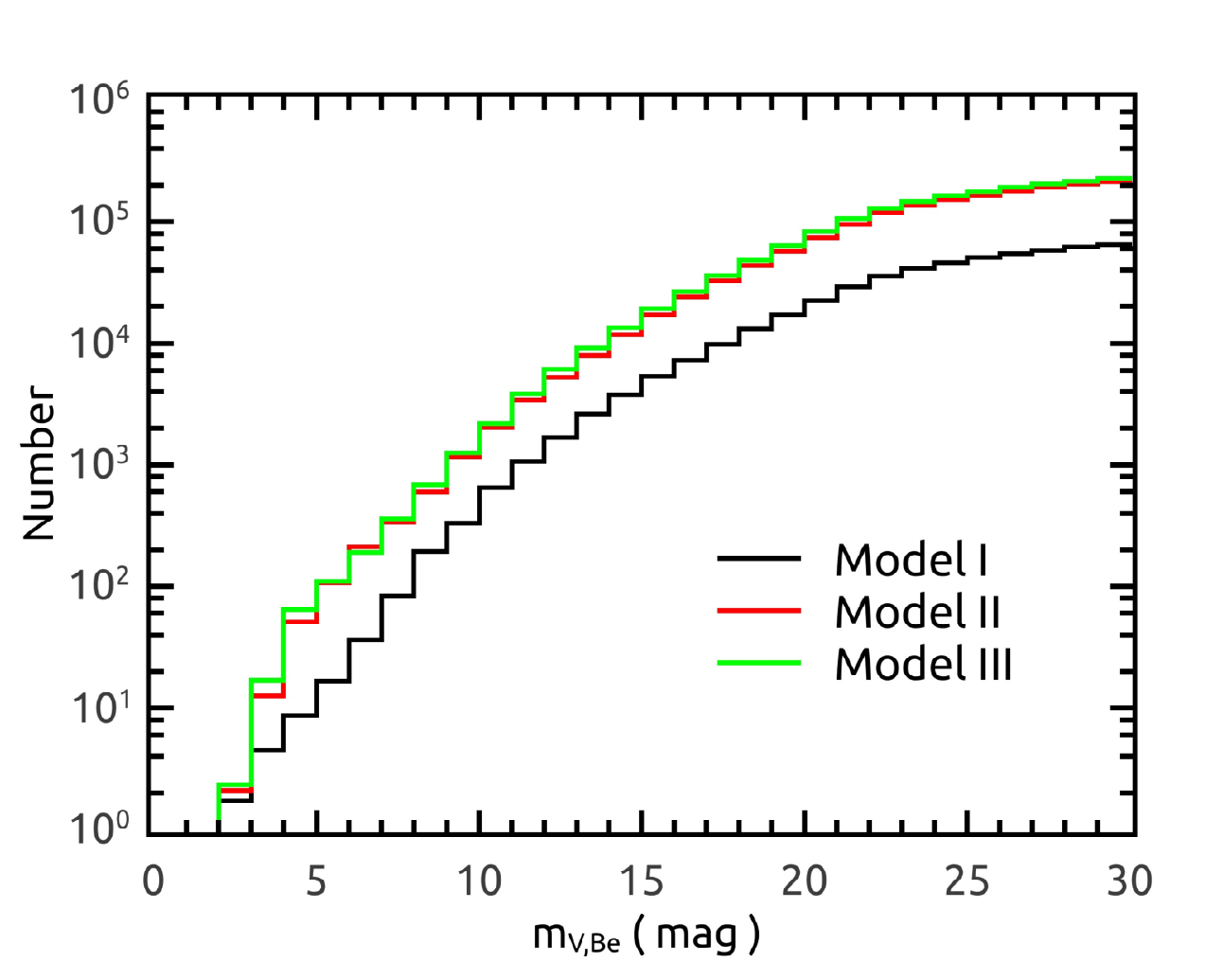}
%\linespread{0.7}
\caption{Accumulated number distributions of Galactic Be$-$He binaries as a function of the V-band apparent magnitude
of the Be stars. The three coloured curves correspond to Models I$-$III. 
   \label{figure2}}
\end{figure}

\begin{figure*}[hbtp]
\centering
\includegraphics[width=1.0\textwidth]{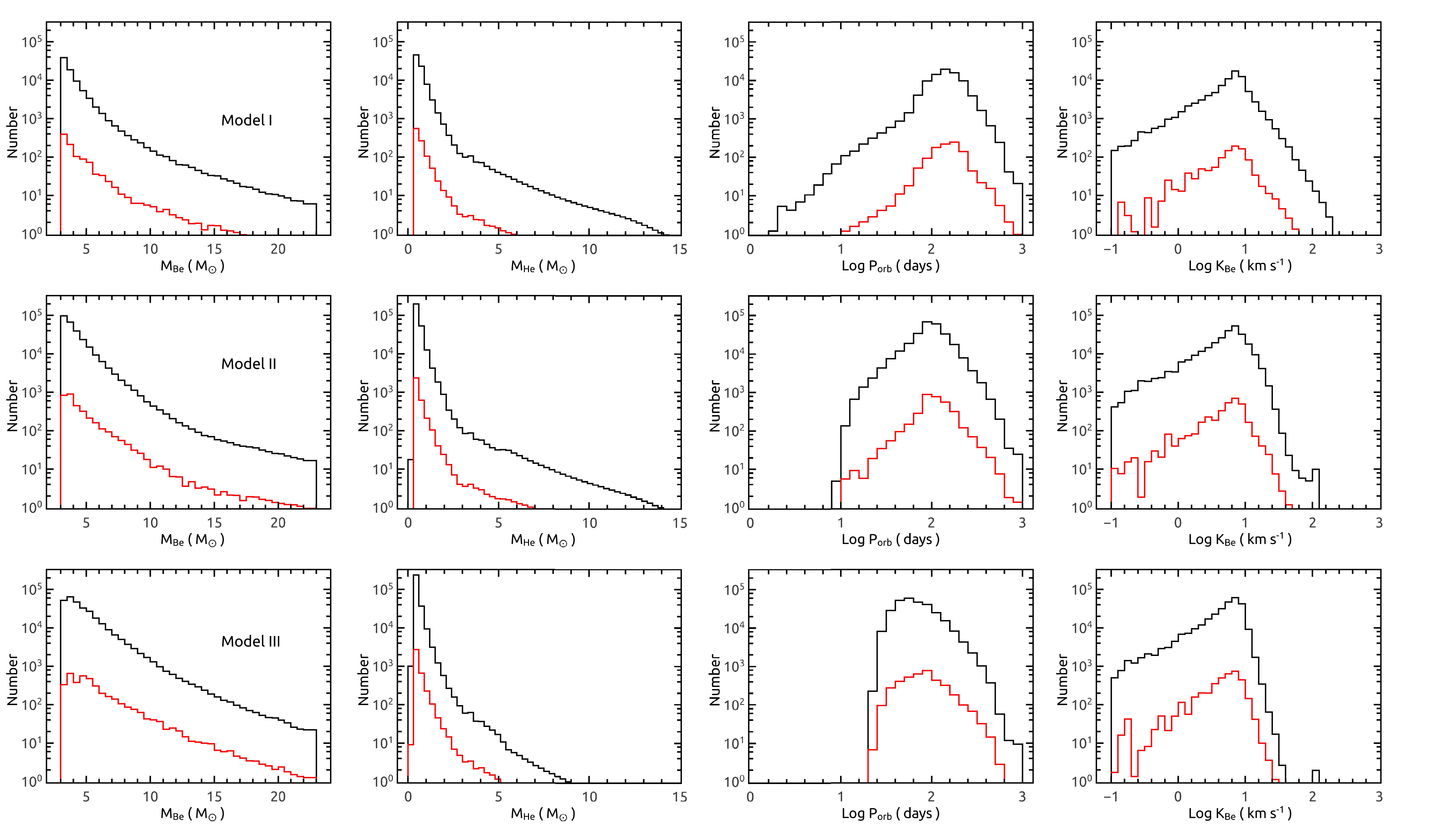}
%\linespread{0.7}
\caption{Calculated number distributions of Galactic Be$-$He binaries as a function of  the Be-star
mass $M_{\rm Be} $, the He-star mass $M_{\rm He} $, the orbital period $P_{\rm orb} $, and 
the radial velocity semi-amplitude $K_{\rm Be} $ of the Be star
in Models I$-$III (from top to bottom). 
The black curves correspond to all Be$-$He binaries, and the red curves correspond to the systems
with Be stars brighter than 12 mag in the V band.
   \label{figure3}}
\end{figure*}

\section{Results}

Figure~1 shows the calculated number distributions of Galactic Be$-$He binaries in the $ M_{\rm Be}-P_{\rm orb} $
and $ M_{\rm Be}-M_{\rm He} $  planes under the assumption of Models I$ - $III.
The top and bottom six panels correspond to the evolution of the progenitor systems experienced the process of 
Case A and Case B mass transfer, respectively. 
Each panel includes a $ 100\times100 $ matrix element for the corresponding binary parameters. The 
color in each pixel reflects the number of Be$-$He binaries in the corresponding matrix element by accumulating 
the product of the birthrates of the binary systems passing through it with the timespan.
The solid diamonds mark the positions of suggested Be$-$He binaries and the purple one corresponds to 
LB-1 if it is a Be$-$He binary.  In Model II, there is a jump at $ M_{\rm He}\sim 2- 5 M_\odot $ in the 
$ M_{\rm Be}-M_{\rm He} $  plane for the systems that experienced Case B mass transfer. The reason is that 
the mass-transfer stability during the primordial binary evolution is sensitive to the orbital periods and the mass ratios, 
leading to the parameter spaces for stable mass transfer be irregular and odd \citep[see Section 2.2 of][]{sl14}.
Our calculations show that the
orbital periods of the binary systems mainly distribute in a range of  $\sim 10$ to a few 100 days. 
Although the mass-transfer efficiencies are significantly different in the three models, they all predict that the 
Be$-$He binaries tend to  have light components due to the IMF. 
%Be$-$He binaries can form from either relatively close primordial binaries undergoing mass transfer when both stars
%are on the main sequence (Case A systems) or relatively wide primordial binaries undergoing mass transfer after the primary
%leaves the main sequence (Case B systems). 
When the mass transfer occurs at the main-sequence stage, the Case A
evolution tends to produce the binaries with relatively large mass ratios
$q$ ($= M_{\rm Be}/M_{\rm He} $), which can extend to $ \gtrsim 5-10$. 
Our calculations show that slightly more ($\sim  50\%$ in Model I, $\sim 70\% $ in Model II, and $\sim 80\% $ in Model III)
Be$-$He binaries are formed via Case B evolution.
In Model III we can well reproduce the observed sample 
except LB-1, while in Models I and II the formation of these systems favours Case A evolution. 
We emphasize that near-conservative mass transfer (i.e., Model III) is unable to form
LB-1 with $q \sim 5 $ \citep{sb20}, because the Be star has accreted too much mass from the progenitor of the He star. 

Fig.~2 presents the cumulative number distributions of Galactic Be$-$He binaries as a function of the V-band apparent
magnitude of the
Be stars in Models I$-$III.  We predict that the total number of Be$-$He binaries in the Milky Way is  
$ \sim 8\times10^{4} $ in Model I, and increases to $ \sim 2. 3 \times 10^{5}$ in Model II and to $ \sim 2.8 \times 10^{5}$ in
Model III \citep[see also][]{sl14}. 
This is because more efficient mass transfer can decrease the lower limit of the primary mass when the minimal mass of Be stars
is fixed to be $ 3M_\odot $, thus increasing the formation rate of Be$-$He binaries
due to the IMF. Considering that LB-1 has the V-band apparent magnitude of $ \sim 12 $ \citep{liu19}, 
we estimate that there are 
$ \sim1100-3800 $ Be$-$He systems with V-band apparent
magnitudes brighter than LB-1. Note that the Be stars are greatly brighter than 
the He stars in the V band for most of Galactic Be$-$He binaries (see Fig.~5), since the He stars are much hotter and 
shine mostly in the UV band. It is expected that a part of mass gainers may spin down and appear
as regular B-type stars due to angular momentum losses via tides and stellar winds. Since the majority of
Galactic Be$-$He systems are expected to contain a late-type Be star, the stellar winds are too weak to spin down 
the Be stars which are able to keep the Be phenomenon in the whole main-sequence stage. The rotation of the mass gainers 
in close binaries tends to synchronize with the orbital motion due to tidal interactions. Since significant orbital angular
momentum loss via the escaped material during the evolution, Model I can produce more binaries with relatively short periods.
We find that the post-mass transfer binaries with a regular B-type star around a He star
are numerous and we find the most in Model I, estimating them to be $ \sim 1.1\times 10^{4} $ in the Galaxy, while in Model
II and III we estimate them to be $ \sim 2.1\times 10^{3} $ and $ \sim 1.1\times 10^{3} $, respectively.
It is indicated that the mass gainers in the majority ($ \gtrsim 90\% $) of post-mass transfer binaries can 
remain to be Be stars in all our models. 
There is a caveat that many binaries with a regular B-type star and a He star may be created via the ejection
of a common envelope, which is not involved in our calculations.

\begin{figure*}[hbtp]
\centering
\includegraphics[width=1.0\textwidth]{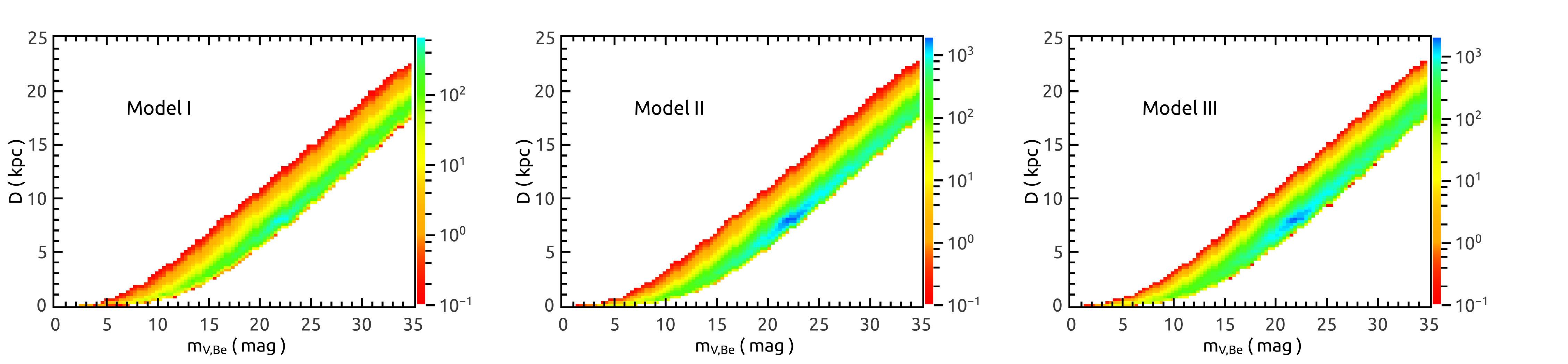}
%\linespread{0.7}
\caption{Calculated number distributions of Galactic Be$-$He binaries in the parameter space of Be-star apparent magnitude vs.
binary distance. The three panels correspond to Models I$ - $III. The colors in each pixel are scaled according to the 
corresponding numbers, only the systems with numbers larger than 0.1 are plotted.
   \label{figure4}}
\end{figure*}

Fig.~ 3 shows the histogram number distributions of the calculated Be$-$He binaries as a function of the Be-star mass,
the He-star mass, the orbital period, and the radial velocity semi-amplitude of the Be star in Models I$-$III (from top to bottom). 
The black curves correspond to all
Be$-$He systems, while the red curves correspond to the Be$-$He systems with Be stars brighter than 12 mag. 
We see that the shapes of the number distribution of Be$-$He binaries in these two cases are similar.  The mass
distribution of the Be stars has a peak at $ \sim 3M_\odot $, and the number of the binary systems rapidly 
decreases with increasing Be-star mass. A similar case can be found for the He-star mass distribution but with a 
peak $ \sim 0.3-0.6M_\odot $.
The peak of the orbital period distribution shifts from $ \sim160 $ days to $ \sim 80 $ days with 
gradually increasing mass-transfer efficiencies from Model I to Model III. In all cases, the radial velocity semi-amplitudes 
of the Be star distribute in the range of $\lesssim 100  \,\rm km\, s^{-1}$ with a peak near $ 8  \,\rm km\, s^{-1}$.

In Fig.~4 we depict predicted number distributions of Galactic Be$-$He binaries in the parameter space of Be-star 
apparent magnitude
vs. binary distance for Models I$-$III. We can see that the binaries with Be stars brighter than 12 mag can be detected within
the distances of $ \lesssim 4$ kpc. A large number of Be$-$He binaries are expected to be located at a distance of $ \sim 5-10 $ kpc, in which the Be stars have the V-band apparent magnitudes of $ \sim20-25 $. 

\section{Observational consequences}

The detection of Be$-$He binaries is subject to serious observational biases \citep{gd17,sg18}. Most of  
systems appear to be single stars. Even for the observed Be$-$He binaries, part of them were suggested to
have atypical parameters. For example, \citet{sg18} pointed out that the stripped He stars in $ \varphi $ Persei and 59 Cyg are remarkably
bright for their current masses. It was explained that these stripped stars reside in the He-shell rather than
He-core burning phase. The detection of such short-lived binaries implies that a large fraction of the systems containing
a less-evolved He star may remain undetected. Also, if LB-1 is a Be$-$He binary, its properties seem to deviate from the typical ones. 
The estimated parameters for the components of LB-1 \citep{sb20} indicate that the stripped star is slightly
colder than the Be star, thus both components can significantly contribute to the optical emission. The relatively 
cold stripped star implies that it is thermally unstable, most likely contracting towards the He main
sequence \citep{sb20}.  This evolutionary phase can last for a duration about $ 1-2 $ orders of magnitude shorter
than the lifetime of a He star \citep{es20}, indicating again that a large number of Be$-$He binaries should exist but be hidden. 

\begin{figure*}[hbtp]
\centering
\includegraphics[width=1.0\textwidth]{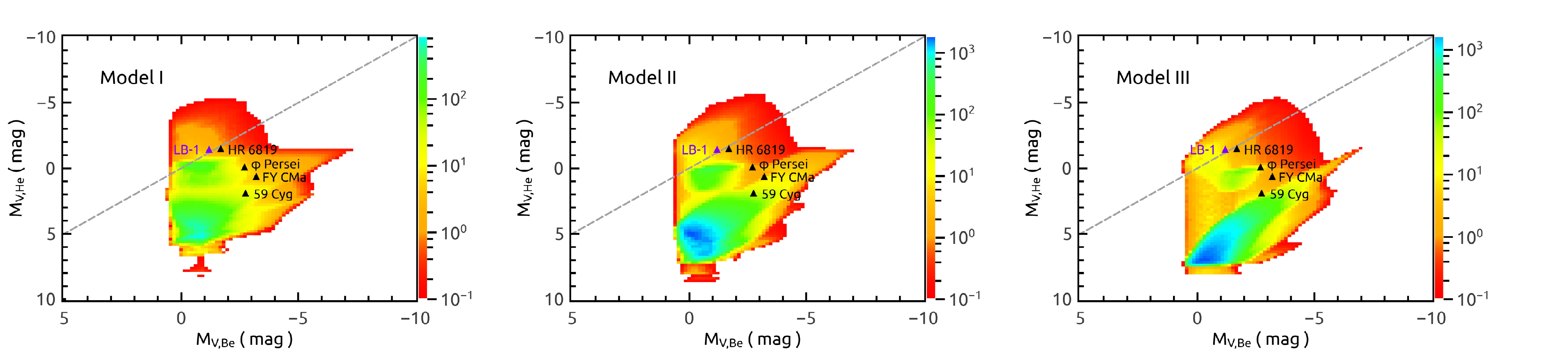}
%\linespread{0.7}
\caption{Calculated number distributions of all Galactic Be$-$He binaries in the $ M_{\rm V,Be}- M_{\rm V,He}$ plane under 
the assumption of  Models I$ - $III (from left to right). The colors in each pixel are scaled according to the 
corresponding numbers, only the systems with numbers larger than 0.1 are
plotted. The positions of suggested Be$-$He systems are plotted with solid triangles$-$the purple one corresponds to 
LB-1 while the black ones correspond to other binaries. The dashed line in each panel corresponds to $ M_{\rm V,Be} = M_{\rm V,He}$. }
   \label{figure4}
\end{figure*}

In Fig.~5, we present the
calculated number distributions of all Galactic Be$-$He binaries in the parameter space of the absolute magnitudes of both
components. The solid triangles mark the positions of observed binaries, for which 
the absolute magnitudes are estimated from the corresponding stellar parameters \citep[data taken from][]{sg18,sb20,bsm20}.
It is obvious that  the calculated Be$-$He binaries mainly distribute in two regions with ($ M_{\rm V, Be}$, $ M_{\rm V, He}$)
clustering around  ($-1 $ mag, 6 mag) and ( $ -1 $ mag,  1 mag). This two-region distribution
is just dependent on whether the He star is in the contraction phase with a bloated envelope\footnote{In order to
identify the binaries with a stripped star being in the contraction phase (labelled as a giant star in the BSE code), we pick 
out all detached Be-star binaries with either a giant star or a He star companion. After using a criterion of 
$ M_{\rm He} < 0.6M_{1}$, we find that the stripped star 
continuously evolves from a giant star to a naked He star for a specific binary.}. Note that
the He stars during the 
He-shell burning phase can also develop a bloated envelope and are relatively bright in the V band, 
the systems with such a He star tend to cover a large region in the  $ M_{\rm V, Be}-M_{\rm V, He}$ plane.
We predict that the numbers of the binary systems with a contracting (expanding) He star
are about  $  1.5\times 10^{4}$ ($  4.9\times 10^{3}$)  in Model I,  $  1.7\times 10^{4}$ ($  1.2\times 10^{4}$) in Model II,  
and $  8\times 10^{3}$ ($  2.7\times 10^{3}$) in Model III.
Compared to the total number of Galactic Be$-$He systems, we can estimate that about $ 4\%-20\% $ of all binaries 
harbour a bloated He star. However, there are only
$ \sim 1\% $ of the Galactic Be$-$He binaries with He stars brighter than Be stars in the V band.

\section{Conclusion}

We use the BPS method to simulate the potential population of Be$-$He binaries in the Milky Way. 
Since the mass-transfer process during the progenitor system evolution is still uncertain, we consider three models 
with significantly different mass-transfer efficiencies to deal with the binary evolution. Comparison 
of the parameter distributions of the calculated Be$-$He binaries with the inferred properties of LB-1 indicates that 
the progenitor system of LB-1 very likely has experienced a non-conservative mass transfer, if it is a Be$-$He system.
Combining with other formation channel(s) of other observed binaries such as $ \varphi $ Persei \citep{p07,sg18},
this indicates that the mass-transfer processes in massive binary evolution are complicated, 
depending on the detailed characteristics of the binaries. We  estimate that 
the total number of Galactic Be$ - $He binaries is of the order $ 10^5$, and $ \sim1100-3800 $ of them are expected to
have the V-band apparent magnitudes brighter than LB-1. If assuming a few percent of such Be$ - $He binaries have a
stripped star being in the contraction phase, there remain at least tens of systems similar to LB-1 to be discovered.

\acknowledgements
We thank the anonymous referee for constructive suggestions that
helped improve this paper.
This work was supported by the Natural Science Foundation 
of China (Nos. 11973026 and 11773015), the Project U1838201 
supported by NSFC and CAS, and the National Program on Key Research and 
Development Project (Grant No. 2016YFA0400803).
%\end{acknowledgements}

\clearpage

\end{document}